\documentclass[fleqn,usenatbib]{mnras}

\usepackage{newtxtext}

\usepackage[T1]{fontenc}

\DeclareRobustCommand{\VAN}[3]{#2}
\let\VANthebibliography\thebibliography
\def\thebibliography{\DeclareRobustCommand{\VAN}[3]{##3}\VANthebibliography}

\usepackage{graphicx}	
\usepackage{amsmath}	
\usepackage{amssymb}	
\usepackage{xcolor}
\definecolor{myblue}{rgb}{0.23, 0.71, 0.83}
\newcommand\zl[1]{{#1}}
\newcommand\blue[1]{{#1}}
\newcommand\cu[1]{{#1}}

\title[Forward-Reverse Shock Model of the GRB Afterglow]{A semi-analytical solution to the forward-reverse shock hydrodynamics of the gamma-ray burst afterglow}

\author[Z. L. Zhang et al.]{
Ze-Lin Zhang,$^{1,2}$
Ruo-Yu Liu,$^{1,2}$\thanks{ryliu@nju.edu.cn}
Jin-Jun Geng,$^{3}$
Xue-Feng Wu$^{3}$\thanks{xfwu@pmo.ac.cn}
and Xiang-Yu Wang$^{1,2}$
\\
\\
$^{1}$School of Astronomy and Space Science, Nanjing University, Xianlin Road 163, Nanjing 210023, China\\
$^{2}$Key laboratory of Modern Astronomy and Astrophysics, Nanjing University, Ministry of Education, Nanjing 210023, China\\
$^{3}$Purple Mountain Observatory, Chinese Academy of Sciences, Nanjing 210023, China
}

\date{Accepted 2022 April 27. Received 2022 April 18; in original form 2022 March 3}

\pubyear{2021}

\begin{document}
\label{firstpage}
\pagerange{\pageref{firstpage}--\pageref{lastpage}}
\maketitle

\begin{abstract}
We extend the standard model of forward-reverse shock (FS-RS) for gamma-ray burst (GRB) afterglow to more general cases. On one hand, we derive the analytical solution to the hydrodynamics of the shocks in two limiting cases, i.e., an ultra-relativistic reverse shock case and a Newtonian reverse shock case. Based on the asymptotic solutions in these two limiting cases, we constitute a semi-analytical  solution for the hydrodynamics of the shocks in the generic case, covering the mildly-relativistic reverse shock case. On the other hand, we derive the evolution of the system taking into account the condition of energy conservation which is not satisfied in the standard FS-RS model. A generic solution of semi-analytical expressions is also given. In both the extended standard FS-RS model  (satisfying pressure balance condition) and the model satisfying energy conservation, we find that the results in the ultra-relativistic reverse shock case and in the early stage of the Newtonian reverse shock case are different from those in the standard FS-RS model by only a factor that close to one while the same initial conditions adopted. However, the asymptotic solutions in the limiting cases are not good approximations to those in the intermediate case. Our semi-analytical results agree well with the numerical results for a large range of model parameters, and hence can be easily employed to diagnose the physical quantities of the GRB shell and circumburst environment.
\end{abstract}

\begin{keywords}
hydrodynamics -- \zl{relativistic processes} -- \cu{shock waves} -- \zl{gamma-ray burst: general} 
\end{keywords}



\section{Introduction} \label{sec:intro}


The standard forward-reverse shock (FS-RS) model for gamma-ray burst (GRB) afterglow was gradually established in 1990s (e.g., \citealt{Rees1992}; \citealt{Katz1994}; \citealt{Sari1995}), and could well interpret the multi-wavelength lightcurves of some GRB afterglows \citep{Meszaros97}. The model considers a cold (the gas pressure is negligible) homogeneous ultra-relativistic shell of GRB eject encounters with interstellar medium (ISM), which generates a forward shock (FS) and a reverse shock (RS) sweeping into ISM and the jet itself respectively (e.g., \citealt{Kobayashi2000}). The entire process is assumed to be adiabatic. Analytical solutions are obtained in two limiting cases, i.e., in an ultra-relativistic reverse shock (RRS) case and a Newtonian reverse shock (NRS) case. After the RS crosses the shell, the shocked ISM and the shocked ejecta are regarded as one body (blastwave region) and their evolution are solved by the Blandford-McKee self-similar solutions \citep{Blandford1976}.

In the standard FS-RS model \citep{Sari1995}, the FS-RS system is divided into four regions by a contact discontinuity (CD) and the two shocks: (1) the unshocked cold ISM; (2) the FS-shocked hot ISM; (3) the RS-shocked hot GRB ejecta and (4) the unshocked cold ejecta, as these regions will be illustrated in Figure \ref{fig:Fig1}. Region 2 (shocked ISM) and region 3 (shocked GRB ejecta) are usually regarded as a whole with the same bulk Lorentz factor $\gamma_2 = \gamma_3$ and the same pressure $p_2=p_3$ or internal energy density $(\hat{\gamma}_{21}-1)e_2=(\hat{\gamma}_{34}-1)e_3$, where $\hat{\gamma}_{21}=\hat{\gamma}_{2}$ and $\hat{\gamma}_{34}=\hat{\gamma}_{3}$ are the adiabatic indices in regions 2 and 3 (hereafter, the quantity of region $i$ will be marked with a subscript ``\,$i$\,''). The evolution of both FS and RS are derived analytically in this scenario for a RRS case and a NRS case. However, in the other word, the derived solutions for the standard FS-RS model works only when the relative Lorentz factor between region 3 and region 4 satisfies $\gamma_{34}-1\ll{1}$ or $\gamma_{34}-1\gg{1}$, which is not the usual situation. Furthermore, the model also requires the circumburst environment to be ISM while \cite{Yi2013} showed that the density profiles of the ambient medium are in between of an ISM and a stellar wind from analyses on early optical afterglows of some GRBs, approximately following a power-law profile of $R^{-k}$ where $k$ ranges from $0.4$ to $1.4$ for studied GRBs. Note that $k=0$ and $k=2$ correspond to the ISM and the stellar wind environment respectively. Even though, as indicated by \cite{Beloborodov2006}, the assumption of pressure balance in the blastwave region (i.e., $p_2=p_3$) violates the energy conservation in an adiabatic system. An accurate treatment of the pressure gradient and Lorentz factor profile in this region can only be achieved by numerical calculation. Many efforts have been made to extend the standard FS-RS model. \blue{\cite{Nakar2004} parametrized the optical emission from RS, with considering an ambient medium to be ISM, to study the time behavior of early optical emissions including a mildly-relativistic RS case (i.e., $\gamma_{34}-1\sim{1}$).} \cu{Furthermore, \cite{Harrison2013} investigated the intermediate reverse shock case based on a better shock approximation and found that the reverse shock emission is much weaker than that estimated by \cite{Nakar2004} in the intermediate regime.} \cite{Dai1998}, \cite{Chevalier1999} and \cite{Wu03} considered a stellar wind environment (region 1) the density of which keeps $n_{1}\propto{R^{-2}}$ with $R$ being the distance to the central engine, and got analytical solutions in the RRS and the NRS cases. 

On the other hand, even though at the CD pressure is balanced, the conditions $\gamma_2 = \gamma_3$ and $(\hat{\gamma}_{21}-1)e_2=(\hat{\gamma}_{34}-1)e_3$ actually make the assumption that, within the blastwave region, both the pressure and the Lorentz factor are homogeneous. In fact, there probably exists a pressure gradient and a non-constant bulk Lorentz factor as well within the blastwave. \cite{Beloborodov2006} relaxed the assumption of $p_2=p_3$ that governs the evolution of $\gamma_3$ (or $\gamma_2$) in the standard FS-RS model, and constructed an independent mechanical model in which they obtained the evolution of the system by solving differential conservation equations of  {energy-momentum and mass flux in the blastwave region.} They introduced a pressure gradient but adopted the  {constant Lorentz factor ($\gamma_2 = \gamma_3\approx\textrm{const}$).} Analytical solutions were founded in the two limiting cases with a power-law density profile of ambient medium, i.e., $n_1\propto{R^{-k}}$. The treatment guarantees energy conservation (\citealt{Uhm2011};\citealt{Uhm2012}) and can more accurately model the emission of the blastwave system. Energy conservation treatment of FS-RS system \citep{Nava13,Geng2014} are also introduced to explain the optical rebrightenings and the polarization evolution during early afterglows of some GRBs (\citealt{Geng2016}; \citealt{Lan2016}).  A review of RS emission in GRBs including an extended model of magnetized shock can be found in \cite{Gao2015}. Recently, \cite{Ai2021} and \cite{Chen2021} considered the energy-conserving model of magnetized FS-RS systems in GRBs \citep{Zhang05}. The importance of the dynamically important magnetic component in shocks have been discussed in \cite{Ai2021} and \cite{Chen2021} with the references included in. Here in this article, we aimed to provide a simple but robust semi-analytic method to the mildly-relativistic FS-RS system for hydrodynamic flow, the case for magneto-hydrodynamic flow will be considered in the future work.

Although one can study the intermediate case or a mildly-relativistic RS case by numerically solving the equations that govern the evolution of the system in any given conditions or by making hydrodynamical simulations \citep{vanEerten10}, simple analytical expressions would be clearly and easily applies to studies of a large sample of GRBs. \blue{In this work, we will adopt the parametrization method used in \citep{Nakar2004} to derive analytical expressions for the evolutions of various physical quantities in the FS-RS system}. A power-law form of the density profile of the ambient medium will be adopted and solutions with arbitrary Lorentz factor of RS will be obtained. The rest part of the paper is arranged as follows. In Section \ref{sec: FS-RS model}, we extend the standard FS-RS model  {(under the condition of pressure balance)} to more general cases first. Based on this, we then focus on the evolution of the system in the energy conservation conditions in Section \ref{sec: conservation}. In both the sections \ref{sec: FS-RS model} and \ref{sec: conservation}, analytical expressions for 
the key physical quantities will be obtained in the two limiting cases (i.e., RRS case and NRS case), and based on which semi-analytical expressions for general solutions within reasonable errors will be constituted. Finally, we give our conclusion in Section \ref{sec: conclusion}.

\section{A general solution to forward-reverse shock model}
\label{sec: FS-RS model}

\begin{figure}
\centering
\includegraphics[width=0.9\columnwidth]{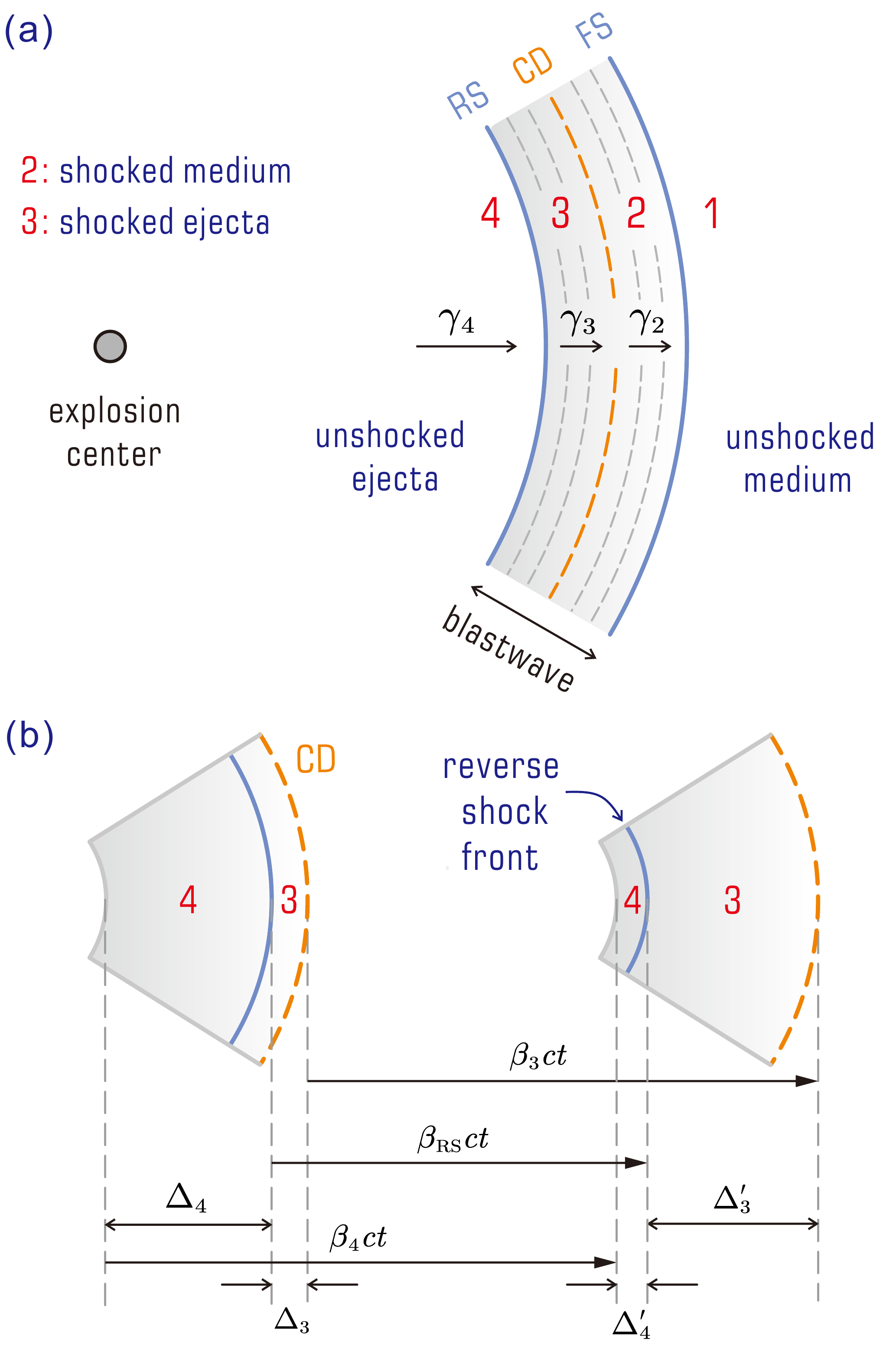}\\ 
\caption{{\bf (a):} Illustrative diagram of four regions  {that are separated by} a spherical blastwave. The system is already described in the sections \ref{sec:intro}. Note that the ambient medium of GRB is not only limited to ISM. {\bf (b):} Illustrative diagram of the GRB ejecta swept up by the RS. $\Delta_4$ is the initial width of the shell and $\Delta_3^{\prime}$ is the width of region 3 after the RS crossing process.} 
\label{fig:Fig1} 
\end{figure}

Let's first denote Lorentz factors $\gamma_i$ and velocities $\beta_i$ of region \textit{i} ($\textit{i}=1,2,3,4$), and thermodynamic quantities of particle number density $n_i$ and internal energy density $e_i$, respectively. Those quantities are related through jump conditions for unmagnetized shocks which read \citep{Blandford1976}
\begin{eqnarray}\label{BM}
  \frac{e_2}{n_2m_pc^2}&=&\gamma_{21}-1,\\
  \frac{n_2}{n_1}&=&\frac{\hat{\gamma}_{2}\gamma_{21}+1}{\hat{\gamma}_{2}-1},\\
  \frac{e_3}{n_3m_pc^2}&=&\gamma_{34}-1,\\
  \frac{n_3}{n_4}&=&\frac{\hat{\gamma}_{3}\gamma_{34}+1}{\hat{\gamma}_{3}-1},
\end{eqnarray}
where $m_p$ is the rest mass of a proton and $c$ is the speed of light. The physical meaning of $\gamma_{34}$ is the bulk Lorentz factor of a downstream (region 3) observer sees a cold upstream (region 4) moving towards the observer and here is regarded as the Lorentz factor of the RS. Similarly, $\gamma_{21}$ can be regarded as the Lorentz factor of the FS. Given $\gamma_4\gg{1}, \gamma_{21}=\gamma_{31}\gg{1}$ which are usually true, we have $\gamma_{34}=\gamma_{43}\simeq(1/2)(\gamma_4/\gamma_3+\gamma_3/\gamma_4)$. The region 2 and region 3 should move with the same Lorentz factor, i.e., $\gamma_{2}=\gamma_{3}$, where we denote $\gamma_{21}=\gamma_{2}$ and $\gamma_{31}=\gamma_{3}$ for simplicity. Otherwise the blastwave region would detach ($\gamma_{3}<\gamma_{2}$) or squeeze ($\gamma_{3}>\gamma_{2}$), as shown in Figure \ref{fig:Fig1}\,(a). $\hat{\gamma}_{2}$ and $\hat{\gamma}_{3}$ are corresponding to the adiabatic indices in regions 2 and 3 where the observer located in, and it can be defined as (e.g., \citealt{Kumar2003}; \citealt{Uhm2011})
\begin{eqnarray}\label{index}
  \hat{\gamma}\simeq\frac{4\bar{\gamma}+1}{3\bar{\gamma}}=
  \begin{cases}
  \displaystyle{5}/{3}, & \bar{\gamma}\sim1~~\,(\textrm{non-relativistic})\cr
  \displaystyle{4}/{3}, & \bar{\gamma}\gg{1}~~(\textrm{relativistic})
  \end{cases},
\end{eqnarray}
 where $\bar{\gamma}$ is the average Lorentz factor of the gas particles. \cu{It
 is worth noting that Equation (\ref{index}) is an assumption and the shocked charged particles (i.e., electrons and protons) are not in equilibrium. As a consequence, they may have different temperatures, especially the temperature of electrons will be much higher if the proportion of electron is not extremely small.}

\begin{figure*}
\centering
\includegraphics[width=1.8\columnwidth]{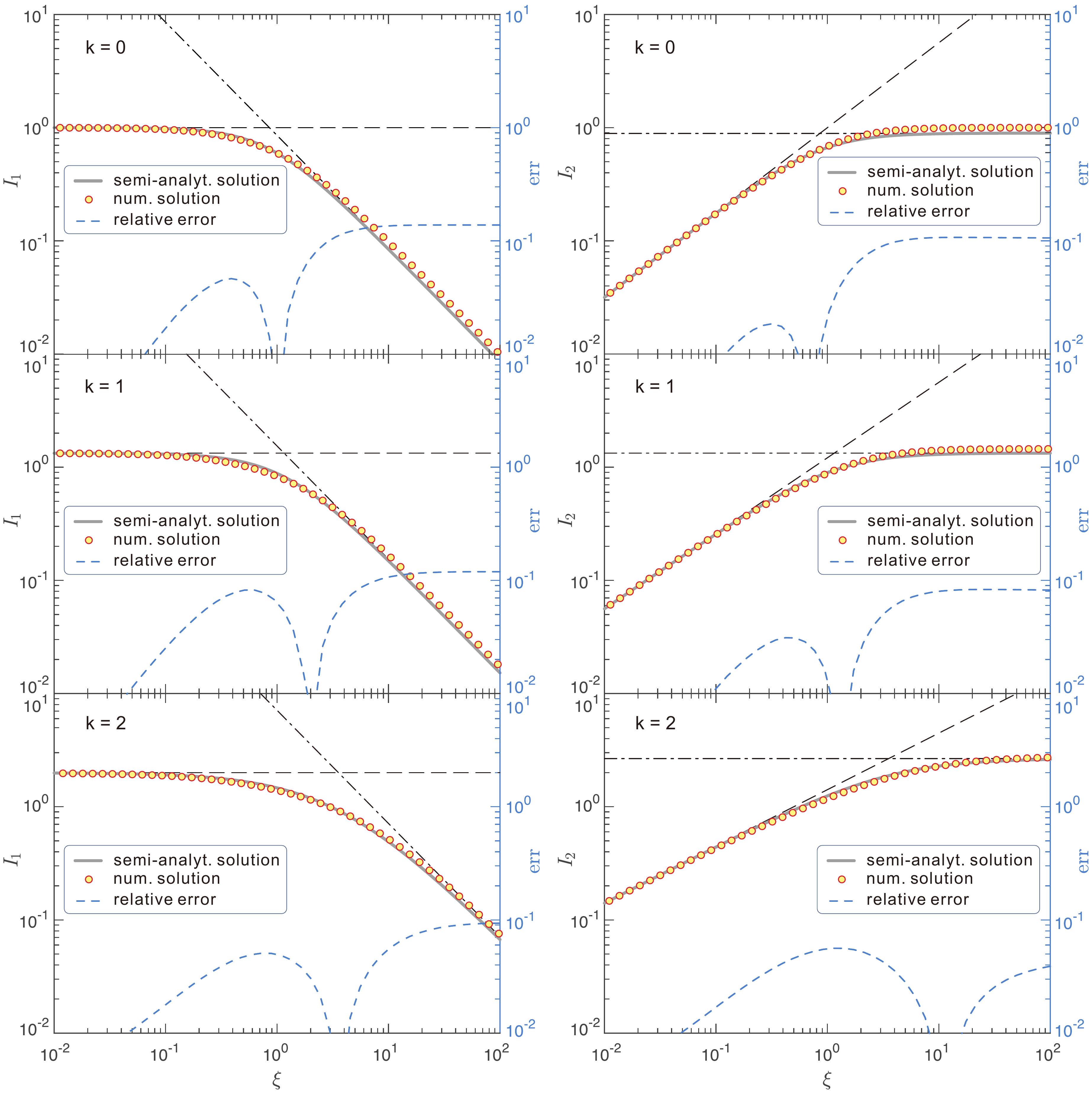}\\ 
\caption{--- $I_1$ and $I_2$ versus $\xi$ under the pressure balance condition. The gray solid lines represent semi-analytical expressions while yellow points exhibit numerical results, the blue dashed lines show the relative error between those two solutions. Black dashed lines and black dash-dotted lines are the analytical solutions in the RRS case and NRS case, respectively.} 
\label{fig:Fig2} 
\end{figure*}

\begin{figure*}
\centering
\includegraphics[width=1.8\columnwidth]{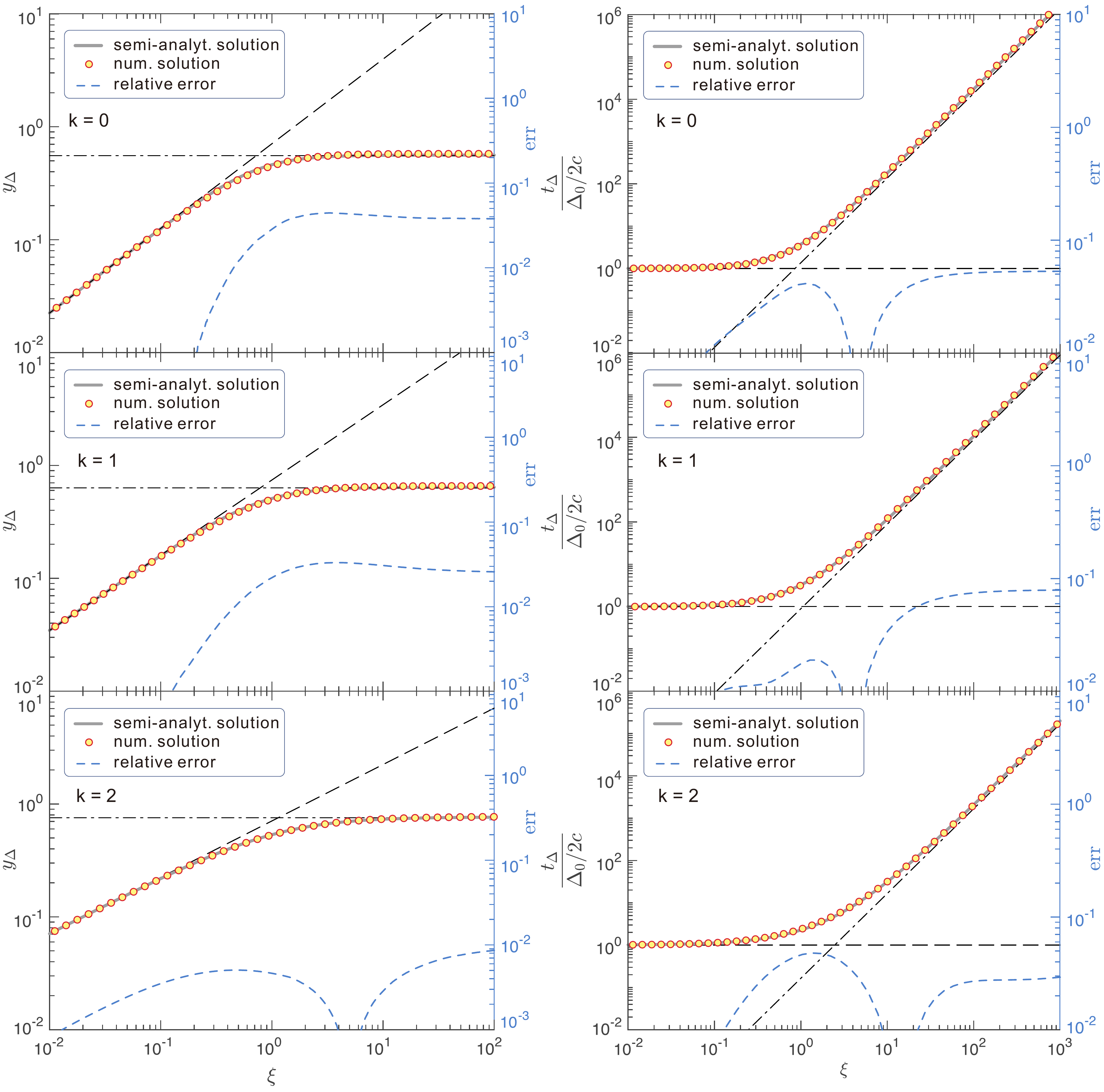}\\ 
\caption{--- The same as Figure \ref{fig:Fig2} but for ${y}_\Delta$ and $\displaystyle{t_\Delta}/[\Delta_0/(2c)]$ versus $\xi$. The asymptotic behavior of the numerical results  {appears} slowly when $\xi\gg{1}$ at the right panels.  {Associate with Figure \ref{fig:Fig2}, we can see that there are some tiny deviations between the numerical solutions and the semi-analytical solutions when $\xi\gg1$, this will be explained in Appendix \ref{sec: numerical}.}} 
\label{fig:Fig3} 
\end{figure*}

As a GRB shell expands to a radius of $R$, the comoving particle number density of the unshocked shell can be given by
\begin{eqnarray}\label{n4}
  n_4=\frac{E}{\displaystyle4\pi{R}^2\gamma_4{m}_pc^2(\gamma_4\Delta)},
\end{eqnarray}
where \cu{$E$ is the initial kinetic energy of the shell}, $\Delta=\Delta_0+R/\gamma_4^2$ is the shell width in the lab frame and $\Delta_0$ is the initial width and $R/\gamma_4^2$ accounts for the expansion of the shell.
Since the width of the shocked region is much smaller than the radius of the shell, we can approximate the density of the ambient medium to be $n_1=AR^{-k}$ if assuming a power-law density profile, $k$ is \zl{a real number} and $0\leq{k}<3$~\cu{(Particularly, $k=0$ and $k=2$ are correspond to homogeneous ISM and typical free stellar wind environment of circumburst medium, respectively)}. 
Here $A$ is a normalization factor. According to \cite{Sari1995}, we define $f\equiv{n_4/n_1}$, with Equation (\ref{n4}) and the definition of $n_1$, we can get
\begin{eqnarray}\label{f}
  f=\frac{l^{3-k}}{(3-k)R^{2-k}\gamma_4^2\Delta},
\end{eqnarray}
where the Sedov length $l$ is defined when the swept ambient medium energy in the shell becomes comparable to the \zl{initial kinetic energy} $E$ of the shell,
\begin{eqnarray}\label{Sedov}
  l=\left[\frac{(3-k)E}{4\pi{A}m_pc^2}\right]^{\frac{1}{3-k}}.
\end{eqnarray}

 {On one hand,} from Equation (1) to Equation (4),
if we take value of $\hat{\gamma}$ to be $4/3$, then we can get the same result as \cite{Sari1995}. Denoting ${y}\equiv{\gamma_3/\gamma_4}$ and adopting the definition of $\hat{\gamma}$ in Equation (\ref{index}), \cu{also note pressure
equality is assumed across the CD, i.e., $p_2=p_3$ and  $(\hat{\gamma}_{2}-1)e_2=(\hat{\gamma}_{3}-1)e_3$, then} we can rewrite and parametrize {Equation (\ref{f})} and get an elegant relation \citep{Zhang2018}:
\begin{eqnarray}\label{ff1}
  f\equiv{\frac{n_4}{n_1}}=\frac{\gamma_2^2-1}{\gamma_{34}^2-1}\simeq\frac{\gamma_3^2}{\gamma_{34}^2-1}=\frac{4\gamma_4^2{y}^4}{\left(1-{y}^2\right)^2}
\end{eqnarray}
or
\begin{eqnarray}\label{ff2}
  \left(1-\frac{4\gamma_4^2}{f}\right){y}^4-2{y}^2+1=0.
\end{eqnarray}
 {This equation is valid during the whole crossing process, and if the value of $f$ is known, then the Lorentz factors of the shocked shell ($\gamma_3={y}\gamma_4$) and the RS ($\gamma_{34}=({y}^2+1)/(2{y})$) can be determined.}
 
On the other hand, to test the validity of our approximation in Equation (\ref{ff1}), let us first compare the results with \cite{Sari1995}. Denoting the width of region 3 and region 4 by, respectively, $\Delta_3$ and $\Delta_4$ at a time $t$ after explosion (in observer's frame) and assume they become $\Delta_3^{\prime}$ and $\Delta_4^{\prime}$ after a short time interval $\textrm{d}t$. So we have
\begin{eqnarray}\label{d3d4}
  \textrm{d}\Delta_3&\equiv&|\Delta_3^{\prime}-\Delta_3|=(\beta_3-\beta_{\rm{RS}})c\textrm{d}t,\\
  \textrm{d}\Delta_4&\equiv&|\Delta_4^{\prime}-\Delta_4|=(\beta_4-\beta_{\rm{RS}})c\textrm{d}t.
\end{eqnarray}
Given that $c\textrm{d}t=\textrm{d}R$, we have $\textrm{d}\Delta_3=(\beta_3-\beta_{\rm{RS}})\textrm{d}R$ and $\textrm{d}\Delta_4=(\beta_4-\beta_{\rm{RS}})\textrm{d}R$. As illustrated in Figure \ref{fig:Fig1}\,(b), from the conservation of baryon number in region 3 and region 4, all the protons that are swept by the RS enter region 3 from region 4, we obtain
\begin{eqnarray}\label{baryon}
  4\pi{R}^2{n}_4(R)\gamma_4\textrm{d}\Delta_4=4\pi{R^{\prime}}^2{n}_3(R^{\prime})\gamma_3\textrm{d}\Delta_3^{\prime}.
\end{eqnarray}
Again, we assume here $R\simeq{R^{\prime}}$ since the width of shocked region is very small comparing to the shock radius, and this leads to
\begin{eqnarray}\label{d4}
  \textrm{d}\Delta_4=\frac{n_3\gamma_3}{n_4\gamma_4}\textrm{d}\Delta_3=\frac{\beta_4-\beta_3}{1-\displaystyle\frac{n_4\gamma_4}{n_3\gamma_3}}\textrm{d}R.
\end{eqnarray}
Considering the jump condition of the RS and bearing in mind the definition of ${y}\,(\equiv\gamma_3/\gamma_4)$, we arrive at (hereafter we simply denote the width of region 4 by $\Delta$)
\begin{eqnarray}\label{dDeltadR}
  \frac{\textrm{d}\Delta}{\textrm{d}R}=\frac{1-{y}^2}{2\gamma_4^2{y}^2}\frac{2{y}^2+2}{2{y}^2+1}.
\end{eqnarray}

Comparing it with $\textrm{d}\Delta/\textrm{d}R=1/(\alpha\gamma_4f^{1/2})$ which is obtained by \cite{Kobayashi1999} and \cite{Kobayashi2000} in the RRS and NRS cases, one can find 
\begin{eqnarray}\label{alpha}
  \alpha=\frac{2{y}^2+1}{2{y}^2+2}.
\end{eqnarray}
The value of $\alpha$ is $1/2$ or $3/4$ ($\approx 0.935\times3/\sqrt{14}$, where $3/\sqrt{14}$ is the result derived by \cite{Sari1995} in the Newtonian limit). If the RS is ultra-relativistic (i.e., ${y}\rightarrow0$) or Newtonian (i.e., ${y}\rightarrow1$), which is almost the same with prefactors in Equations (6) and (8) of \cite{Sari1995} respectively. One can see that under the case of NRS, the difference between our result and \cite{Sari1995} is only a multiple (0.935) close to $1$.

\begin{table*}
	\centering
	\caption{Semi-analytical expressions for $I_1,\,I_2,\,{y}_\Delta$ and $t_\Delta$ under the pressure balance condition.}
	\begin{tabular}{ccllcl} 
		\hline\hline
		& Expression & $f_1$ & $f_2$ & $s$ & Error\\
		\hline
 \textsc{$I_1$} & $\displaystyle\left(f_1^{-s}+f_2^{-s}\right)^{-1/s}$ & $\displaystyle\frac{4}{4-k}$ & $\displaystyle\left[\frac{8}{3(3-k)}\right]^{\frac{4-k}{3-k}}\xi^{-1}$ &  {$1.6+0.05k-0.2k^2$} &  {$<14\,\%$} \\
 
 \textsc{$I_2$} & $\displaystyle(f_1^{-s}+f_2^{-s})^{-1/s}$ & $\displaystyle\left(\frac{4\,\xi}{4-k}\right)^{\frac{3-k}{4-k}}$ & $\displaystyle\frac{8}{3(3-k)}$ &  {2} &  {$<12\,\%$} \\
 
 \textsc{${y}_\Delta$} & $\displaystyle(f_1^{-s}+f_2^{-s})^{-1/s}$ & $\displaystyle\left[\frac{1}{4}\left(\frac{4}{4-k}\right)^{\frac{2(2-k)}{4-k}}\right]^{\frac{1}{4}}\xi^{\frac{3-k}{4-k}}$ & $\displaystyle\frac{2}{\sqrt{13-3k}}$ & $2.02+2.4k-3.63k^2+1.23k^3$ & $<5\%$ \\
 
 \textsc{$\displaystyle\frac{t_\Delta}{{\Delta_0}/{2c}}$} & $\displaystyle(f_1^{s}+f_2^{s})^{1/s}$ & $1$ & $\displaystyle{{}^{\textrm{a}}\mathcal{Q}(k)\left[\frac{8}{3(3-k)}\right]^{-\frac{2}{3-k}}\xi^{2}}$ & $0.68-0.058k-0.01k^2-0.007k^3$ & $<8\%$ \\
		\hline\hline
\end{tabular}
\label{tab:sims}
\vspace{0.1cm}
\newline
$[{\textrm{a}}]$: $\mathcal{Q}(k)$ is represented by Equation (\ref{Q1}). ~~~~~~~~~~~~~~~~~~~~~~~~~~~~~~~~~~~~~~~~~~~~~~~~~~~~~~~~~~~~~~~~~~~~~~~~~~~~~~~~~~~~~~~~~~~~~~~~~~~~~~~~~~~~~~~~~~~~~~~~~~~~~~~~~~~~~~~~~~~~~~~~~~~~~~~~~~~~~~~
\end{table*}

 Now, let's consider the general cases of FS-RS system which are not only satisfying the two limiting cases mentioned above, but also including the intermediate cases. \cu{Here we adopt the parametrization method used in \cite{Nakar2004}, while our treatment of the number of particles in region 4 is different from \cite{Nakar2004} and \cite{Harrison2013}.} In order to construct semi-analytical expressions including the case of mildly-relativistic RS, we first set up the connection between the evolution of FS-RS system in two limiting cases with the number of particles in region 4. As RS crossing the shell of GRB ejecta, particles in the region 4 propagating into the region 3 and the kinetic energy of the ejecta is converted into the thermal energy of shocked particles. At the crossing time, all the particles in the region 4 are swept by the RS, so we have the relation in particles' comoving frame
\begin{eqnarray}\label{relation}
\begin{aligned}
N_{4} &=\int_{0}^{R_{\Delta}} 4 \pi R^{2} n_{4} \gamma_{4} \mathrm{~d} \Delta \\
&=\int_{0}^{R_{\Delta}} \frac{4 \pi A}{\alpha} R^{2-k} f^{1 / 2} \mathrm{~d} R,
\end{aligned}
\end{eqnarray}
where $R_{\Delta}$ is defined as the RS crossing radius. Denoting ${x}\equiv{R}/{R_{\Delta}}$, and using Equations (\ref{f}) and (\ref{Sedov}), the above equation can be written into
\begin{eqnarray}\label{int1int2}
\begin{aligned}
N_4&\equiv \frac{E}{\gamma_4m_pc^2}\\
&=\begin{cases}
  \displaystyle\frac{4\pi{A}l^{\frac{3-k}{2}}R_{\Delta}^{\frac{4-k}{2}}}{(3-k)^{\frac{1}{2}}\Delta_0^{\frac{1}{2}}\gamma_4}\int_0^{1}\frac{{x}^{\frac{2-k}{2}}}{\alpha\left(1+\displaystyle{x}\frac{R_{\Delta}}{\gamma_4^2\Delta_0}\right)^{\frac{1}{2}}}\textrm{d}{x},\\ \\
  \displaystyle\frac{4\pi{A}(lR_{\Delta})^{\frac{3-k}{2}}}{(3-k)^{\frac{1}{2}}}\displaystyle\int_0^{1}\frac{{x}^{\frac{1-k}{2}}}{\alpha\left(1+\displaystyle\frac{1}{{x}}\frac{{\gamma_4^2\Delta_0}}{{R_\Delta}}\right)^{\frac{1}{2}}}\textrm{d}{x}.
  \end{cases}
\end{aligned}
\end{eqnarray}
 The upper and lower formulae are equivalent to each other, but the upper one is more convenient to use in a thick-shell case (i.e., $\Delta_0\gg R_{\Delta}/\gamma_4^2$), which usually corresponds to a RRS case because given the total particle number in the shell, a thicker shell leads to a lower particle density in the shell, and hence can form a stronger RS when the shell collides with the ambient medium. In contrast, the lower equation is better for a thin-shell case (i.e., $\Delta_0\ll R_{\Delta}/\gamma_4^2$), which often corresponds to a NRS case. Let us denote the integral part in the upper and the lower formulae of Equation (\ref{int1int2}) by $I_1$ and $I_2$, and follow \cite{Sari1995} to define a dimensionless quantity,
\begin{eqnarray}\label{ksi}
  \xi \equiv  (3-k)^{-\frac{1}{2(3-k)}}\left(\frac{l}{\Delta_0}\right)^{\frac{1}{2}}\gamma_4^{-\frac{4-k}{3-k}},
\end{eqnarray}
which only depends on parameters of GRB ejecta and ambient environments, {as mentioned in Equation (\ref{f}) and Equation (\ref{Sedov})}. \zl{Note that $\xi\ll{1}$ gives rise to a thick-shell case while $\xi\gg{1}$ results in a thin-shell case.} Then we can find a key relation that
\begin{eqnarray}\label{Iksi}
  \frac{R_{\Delta}}{\gamma_4^2\Delta_0} = I_1^{-\frac{2}{4-k}}\xi^{\frac{2(3-k)}{4-k}}=I_2^{-\frac{2}{3-k}}\xi^{2}.
\end{eqnarray}
If substituting this relation into the expressions of $I_1$ and $I_2$, one will obtain
\begin{eqnarray}\label{Pi1}
  I_1 = \int_{0}^{1}\frac{1}{\alpha(y(x))}{x}^{\frac{2-k}{2}}\left[1+I_1^{-\frac{2}{4-k}}\xi^{\frac{2(3-k)}{4-k}}{x}\right]^{-\frac{1}{2}}\textrm{d}{x}~~
\end{eqnarray}
and
\begin{eqnarray}\label{Pi2}
  I_2 = \int_{0}^{1}\frac{1}{\alpha(y(x))}{x}^{\frac{1-k}{2}}\left[1+I_2^{\frac{2}{3-k}}\xi^{-2}{x}^{-1}\right]^{-\frac{1}{2}}\textrm{d}{x}.~~
\end{eqnarray}
 {Here, $y$ is the functional form of $x$ and it is given by $y=y(x)$}. It is not easy to find analytical expressions for $I_1$ and $I_2$ from above integrals, but if the RS is ultra-relativistic ($\xi\ll{1}$) or Newtonian ($\xi\gg{1}$) we can easily reach
\begin{eqnarray}\label{Pi11}
I_1=\begin{cases}
  \displaystyle\frac{4}{4-k}, & \xi\ll{1},\cr\cr
  \displaystyle\left[\frac{8}{3(3-k)}\right]^{\frac{4-k}{3-k}}\xi^{-1}, & \xi\gg{1}.
  \end{cases}
\end{eqnarray}
and
\begin{eqnarray}\label{Pi22}
I_2=\begin{cases}
  \displaystyle\left(\frac{4\,\xi}{4-k}\right)^{\frac{3-k}{4-k}}, ~~~~~~& \xi\ll{1},\cr\cr
  \displaystyle\frac{8}{3(3-k)}, ~~~~~~& \xi\gg{1}.
  \end{cases}
\end{eqnarray}

 The values of $I_1$ and $I_2$ at $\xi\sim1$ should be in between the values in the two limiting cases. Take $I_1$ for instance. If we denote the solution for $I_1$ when $\xi\ll{1}$ by $f_1$ and that when $\xi\gg{1}$ by $f_2$, a semi-analytical expression for $I_1$ in the whole considered range of $\xi$ can be constituted by
 {\begin{eqnarray}\label{f1f2}
I_1=\displaystyle\left(f_1^{-s}+f_2^{-s}\right)^{-\frac{1}{s}}.
\end{eqnarray}}
This expression ensures the behavior of $I_1$ depends on $f_1$ when $f_1\ll{f_2}$ or $\xi\ll{1}$, and depends on $f_{2}$ when $f_1\gg{f_2}$ which happens in the case of $\xi\gg{1}$. The parameter $s$ determines the sharpness of the transition from $f_1$ to $f_2$. Comparing with the numerical results, we find that the expression can give a good description of $I_1$ by taking
 {\begin{eqnarray}\label{s0}
{s(k)}=-0.2k^2+0.05k+1.6
\end{eqnarray}
for $k$ from $0$ to $2$, while the relative error of the semi-analytical expression to the numerical one can be controlled within $14\,\%$. A similar semi-analytical expression can be also constituted for $I_2$ with relative error less than $12\,\%$ when taking $s=2$ for $k$ from $0$ to $2$.} In Figure \ref{fig:Fig2}, we present both the semi-analytical and numerical results of $I_1$ and $I_2$ versus $\xi$ as well as the relative errors in $k=0, 1, 2$ cases.  {The main deviation of numerical solution from the analytical solution appears in the limit $\xi \gg 1$. This is because the shocked GRB shell does get decelerated during the RS crossing in the thin-shell case and hence the value of $\alpha$ (see Equation (\ref{alpha})) can not really maintain $3/4$ all the time, as is shown in Appendix \ref{sec: numerical}.} Nevertheless, the resultant $\sim 10\%$ error is acceptable. We summarize the semi-analytical expressions of $I_1,\,I_2$ in Table \ref{tab:sims}.

Once we have the expressions of $I_1$ and $I_2$, it is possible to obtain the crossing radius $R_{\Delta}$ through Equation (\ref{Iksi}), and then we can know the value of $f$ at that time. Note that Equation (\ref{ff2}) holds for the whole crossing process. In a thick-shell case ($\xi\ll{1}$), we expect ${y}\ll{1}$ and $\gamma_4^2/f\gg{1}$. So Equation (\ref{ff2}) can be reduced to ${y}=[f/(4\gamma_4^2)]^{1/4}$. For $\xi\ll{1}$, we have $\Delta_0\gg{R}_{\Delta}/\gamma_4^2$ and $I_1=4/(4-k)$, utilizing Equation (\ref{f}) and $R={x}{R}_{\Delta}$, we have
\begin{eqnarray}\label{chi10}
{y}=\left[\frac{1}{4}\left(\frac{4}{4-k}\right)^{\frac{2(2-k)}{4-k}}\right]^{\frac{1}{4}}\xi^{\frac{3-k}{4-k}}{x}^{\frac{k-2}{4}}.
\end{eqnarray}
At the crossing radius, we can simply take ${x}=1$ in the above equation to get
\begin{eqnarray}\label{chi11}
{y}_\Delta=\left[\frac{1}{4}\left(\frac{4}{4-k}\right)^{\frac{2(2-k)}{4-k}}\right]^{\frac{1}{4}}\xi^{\frac{3-k}{4-k}}.
\end{eqnarray}
 {In addition}, none of the terms in Equation (\ref{ff2}) can be neglected in a thin-shell case ($\xi\gg{1}$). Given
\begin{eqnarray}\label{ff3}
f\simeq\displaystyle\frac{l^{3-k}}{(3-k)({x} R_\Delta)^{3-k}}
\end{eqnarray}
and $I_2=8/[3(3-k)]$, we find $f/(4\gamma_4^2)=[3(3-k)/4]^{-2}{x}^{k-3}$ and Equation (\ref{ff2}) can be reduced to
\begin{eqnarray}\label{ff22}
  \left(1-\left[\frac{3(3-k)}{4}\right]^2{x}^{3-k}\right){y}^4-2{y}^2+1=0.
\end{eqnarray}
Assuming $x=1$ and picking the positive real solution of above equation, we get 
\begin{eqnarray}\label{ydelta1}
y_\Delta=\frac{2}{\sqrt{13-3k}}.
\end{eqnarray}
Again, we denote Equation (\ref{chi11}) by $f_1$ and Equation (\ref{ydelta1}) by $f_2$, and constitute a semi-analytical solution to ${y}_\Delta$ for an arbitrary $\xi$ with the expression ${y}_\Delta=(f_1^{-s}+f_2^{-s})^{-1/s}$, where
\begin{eqnarray}\label{ss}
s(k)=1.23k^3-3.63k^2+2.4k+2.02
\end{eqnarray}
and the corresponding relative error are described in Table \ref{tab:sims}. Equation (\ref{ydelta1}) indicates that even if in the thin-shell case, the shocked region are decelerated, although may not be significantly, and the RS can convert a non-negligible fraction of the kinetic energy of the GRB ejecta to the thermal energy of shocked particles. In contrast, the RS is often thought to dissipate tiny kinetic energy in the standard FS-RS model and the shocked region still maintain the initial Lorentz factor. This would result in a difference in the estimation of Lorentz factor of the RS in the thin-shell case, and may further lead to differences in predicted light curve of the early afterglow.

\begin{figure}
\centering
\includegraphics[width=0.92\columnwidth]{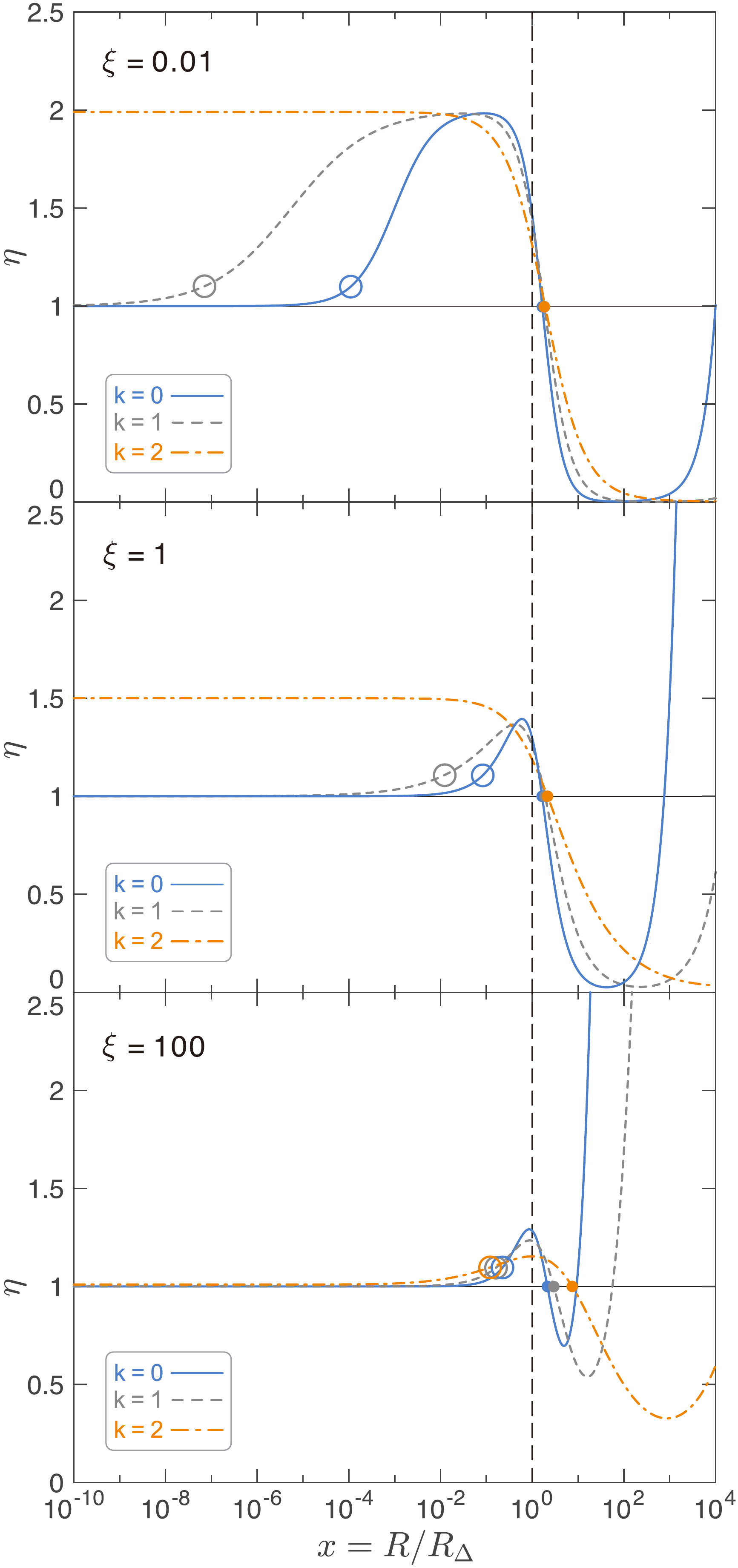}\\ 
\caption{--- A test of energy conservation under the pressure balance condition.  \zl{Lines from top panel to bottom panel (from RRS case to intermediate case, then to NRS case) represent the ratio $\eta$ of total energy of cold gases in GRB shell and ISM to the total energy of these gases after being shocked by RS and FS respectively. Different ambient environments of $k=0, 1, 2$ are represented by blue solid, gray dashed and orange dash-dotted lines, separately. Open circles of different colors show the locations where $\eta$ first deviates (refer to the relative error of $10\,\%$ of $\eta=1$) from unity among the three cases, and points of different colors show the location where $\eta$ goes back to unity again following the shock crossing. The vertical long black dashed line show the location of crossing radius.}} 
\label{fig:Fig4} 
\end{figure}

\begin{figure*}
\centering
\includegraphics[width=2\columnwidth]{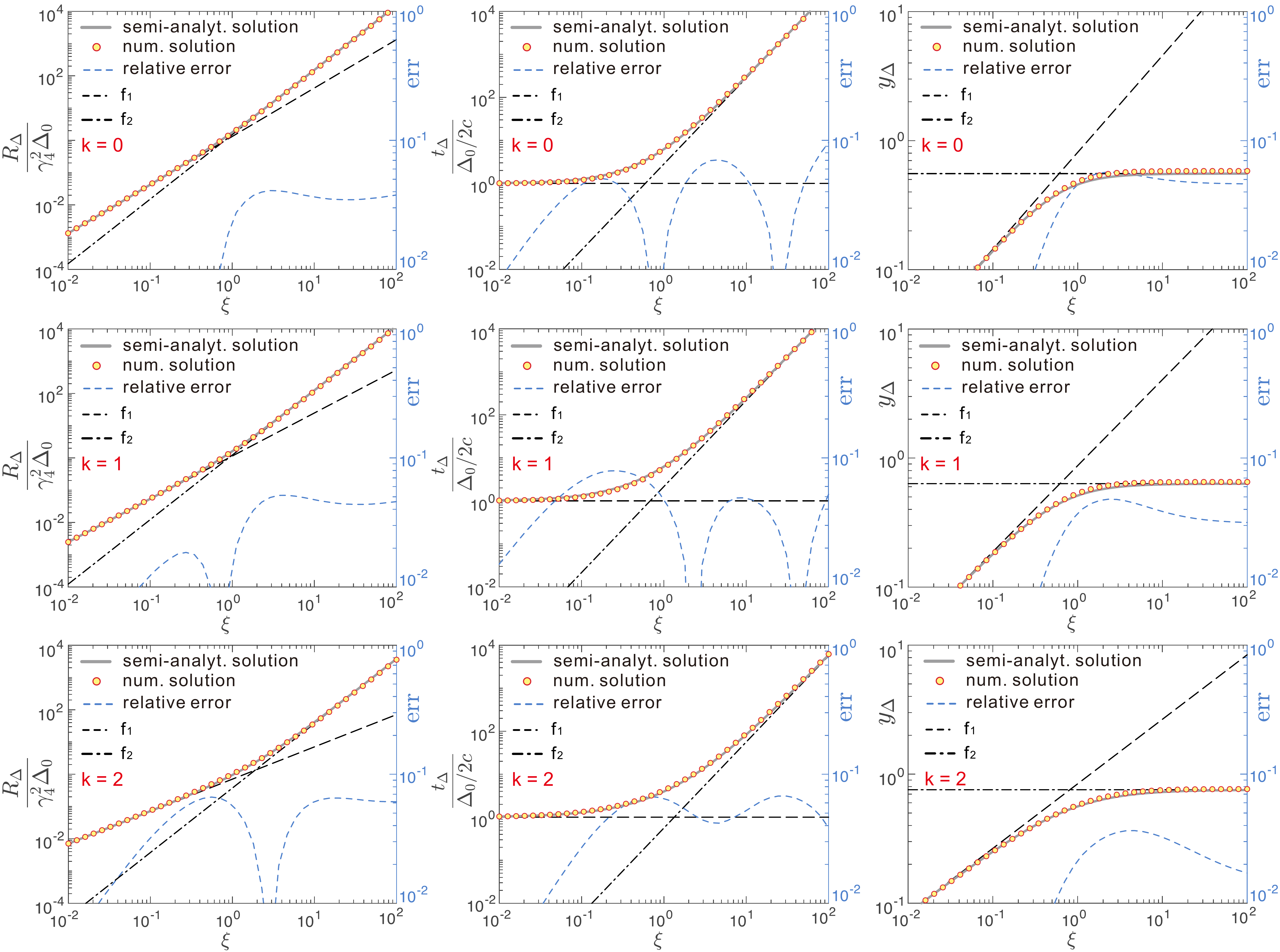}\\ 
\caption{--- $R_\Delta/(\gamma_4^2\Delta_0)$, $\displaystyle{t_\Delta}/[\Delta_0/(2c)]$ and ${y}_\Delta$ versus $\xi$ under the condition of energy conservation. Gray solid lines and yellow points represent semi-analytical expressions and numerical results respectively. Blue dashed lines show the relative error between the semi-analytical expressions and numerical results. Black dashed line and black dash-dotted lines are the analytical solutions in the RRS case and NRS case, respectively.} 
\label{fig:Fig5} 
\end{figure*}

The RS crossing time $t_\Delta$ can be obtained in a similar way. Given that $\textrm{d}t=\textrm{d}R/(2\gamma_3^2c)=R_\Delta/(2\gamma_4^2c)\textrm{d}{x}/{y}^2$, we have the dimensionless parameter 
\begin{eqnarray}\label{Q0}
\mathcal{Q}\equiv\frac{2\gamma_4^2ct_{\Delta}}{R_\Delta}=\int_0^1\frac{\textrm{d}{x}}{{y}^2}.
\end{eqnarray}
Substituting Equation (\ref{chi10}) into $\mathcal{Q}$, we have $\mathcal{Q}=[4/(4-k)]^{2/(4-k)}\xi^{-2(3-k)/(4-k)}$ in the thick-shell case, and this gives $t_\Delta=\Delta_0/(2c)$. Note that the crossing time obtained by \cite{Sari1995} in the thick-shell case is $\Delta_0/c$. That's because they employed the relation $\textrm{d}t=\textrm{d}R/(\gamma_3^2c)$ instead of $\textrm{d}t=\textrm{d}R/(2\gamma_3^2c)$. In the thin-shell case, a relation between $x$ and $y$ is hard to be obtained analytically through Equation (\ref{ff22}), but this equation implies that $\mathcal{Q}$ only depends on $k$. By fitting the numerical results, we find
\begin{eqnarray}\label{Q1}
\mathcal{Q}(k) = 0.02k^2-0.085k+1.26
\end{eqnarray}
and
\begin{eqnarray}\label{TDT}
t_{\Delta} = \mathcal{Q}(k)\frac{\Delta_0}{2c}\left[\frac{8}{3(3-k)}\right]^{-\frac{2}{3-k}}\xi^{2}.
\end{eqnarray}
Note that the value of $t_{\Delta}$ increases with $\xi$, which is unlike the case in $I_1$ and $I_2$ and $y_{\Delta}$, so we use another formula, say, $f=(f_1^s+f_2^s)^{1/s}$, to constitute the semi-analytical expression of $t_\Delta$ with
\begin{eqnarray}\label{sss}
s(k) = -0.007k^3-0.01k^2-0.058k+0.68.
\end{eqnarray}
Figure \ref{fig:Fig3} shows the changes of ${y}_\Delta$ and $t_\Delta/[\Delta_0/(2c)]$ with $\xi$ in $k=0, 1, 2$ case. We also summarize the semi-analytical expressions of ${y}_\Delta$ and $t_\Delta$ in Table \ref{tab:sims}. From Figure \ref{fig:Fig2} and Figure \ref{fig:Fig3}, one can find that analytical solutions in the two limiting cases can not be generalized to the intermediate case with $0.1\lesssim\xi\lesssim10$ with a satisfactory accuracy,  {especially in the $k=0$ case}. Thus, our semi-analytical expressions will help to get a more accurate estimation of those physical quantities.

\section{Energy conservation solution}
\label{sec: conservation}

\begin{figure*}
\centering
\includegraphics[width=1.8\columnwidth]{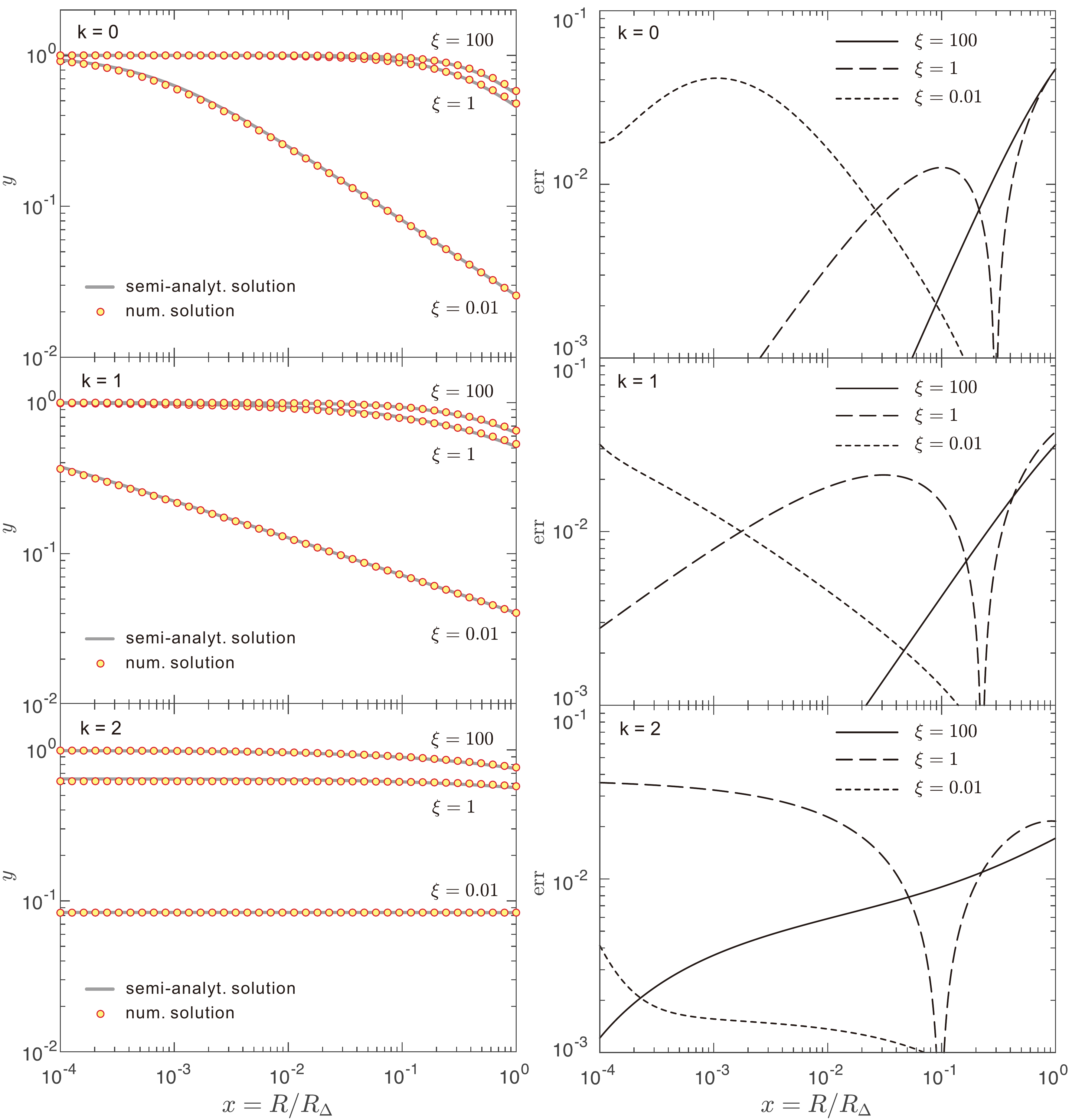}\\ 
\caption{--- Evolution of the Lorentz factor of the shocked gas with the shock radius in different conditions. The solid gray lines exhibit the semi-analytical expressions and yellow points show the numerical results. The black short dashed lines, long dashed lines and solid lines correspond to the relative errors of $\xi=0.01,1$ and $100$ in different cases ($k=0, 1, 2$). \cu{The evolution of $y$ versus $\xi$ and $x$ can be found in Figure \ref{fig:Fig7} and \ref{fig:Appendix2}, One can find the corresponding [R,G,B] color between the contour plot and the colorbar to settle down the value of $y$.
}} 
\label{fig:Fig6} 
\end{figure*}

In the previous section, we assume a uniform pressure and Lorentz factor in the region 2 and the region 3, depending on the standard FS-RS model. However, this condition is not met in reality, since it violates energy conservation of the system (e.g., \citealt{Huang99}; \citealt{Peer12}), which reads
\begin{eqnarray}\label{energy1}
m_2c^2+\gamma_4m_3c^2=\gamma_3^2m_2c^2+\gamma_3\gamma_{34}m_3c^2,
\end{eqnarray}
where $m_2=4\pi m_p{A}R^{3-k}/(3-k)$ is the mass of swept gas by the FS. The two terms in right-hand side are the total energy (including thermal energy, kinetic energy and rest energy) in the region 2 and the region 3 respectively, while the terms in left-hand side show the total energy (kinetic energy and rest energy) before they are swept, respectively, by the FS and the RS. Figure \ref{fig:Fig4} presents \zl{the ratio $\eta$ of total energy $E_{\rm i}$ of cold gases in GRB shell and ISM to the total energy $E_{\rm f}$ of these gases after being shocked by RS and FS respectively}, with the assumption of the pressure balance. \cu{Then we have $\eta=E_{\rm i}/E_{\rm f}$, where $E_{\rm i}=m_2c^2+\gamma_4m_3c^2$ and $E_{\rm f}=\gamma_3^2m_2c^2+\gamma_3\gamma_{34}m_3c^2$.} One can see that the values deviate from unity \zl{significantly especially in the cases of RRS and intermediate ($\xi=0.01$ and $\xi=1$), implying that the total energy is no more conserved, as shown in the top and middle panels of Figure \ref{fig:Fig4}. The crossing radius $R_{\Delta}$ is calculated by Equation (\ref{Iksi}) and the second line of Table \ref{tab:sims}.} In the following part of this section, we aim to derive analytical solutions based on the condition of energy conservation.

\subsection{\zl{Theoretical treatment}}
\label{subsec: Theoretical}
we still define ${y}\equiv\gamma_3/\gamma_4$. Given that $\gamma_3,\,\gamma_4\gg{1}$, the energy conservation equation can be written to be
\begin{eqnarray}\label{energy2}
(1-{y}^2)m_3c^2=2\gamma_4{y}^2m_2c^2.
\end{eqnarray}
The mass in the region 3 increases and the width of the region 4 is shortened as the RS sweeps into the unshocked ejecta. This process in the lab frame can be depicted by
\begin{eqnarray}\label{m3}
\textrm{d}m_3=4\pi{R}^2m_p(\gamma_4n_4)\textrm{d}\Delta.
\end{eqnarray}


Substituting Equation (\ref{energy2}) to Equation (\ref{dDeltadR}), we have
\begin{eqnarray}\label{dDelta2}
\textrm{d}\Delta=\frac{m_2}{\gamma_4m_3}\frac{2{y}^2+2}{2{y}^2+1}\textrm{d}R,
\end{eqnarray}
and substitute it into Equation (\ref{m3}), we obtain
\begin{eqnarray}\label{m31}
m_3\textrm{d}m_3=4\pi{R}^2n_4m_pm_2\frac{2{y}^2+2}{2{y}^2+1}\textrm{d}R.
\end{eqnarray}
In the thick-shell case or the RRS case, ${y}\rightarrow0$ in most of the time of the crossing process, so Equation (\ref{m31}) can be written into
\begin{eqnarray}\label{m32}
m_3\textrm{d}m_3=\frac{8\pi EAm_p}{\displaystyle(3-k)\gamma_4^2\Delta{c}^2}R^{3-k}\textrm{d}R.
\end{eqnarray}

We integrate this equation to get the evolution of $m_3$ with $R$,
\begin{eqnarray}\label{F1}
m_3=\displaystyle\sqrt{\frac{16\pi EAm_pR^{4-k}{}_2{F}_1\left(1,4-k;5-k;\displaystyle-\frac{R}{\gamma_4^2\Delta_0}\right)}{\displaystyle(3-k)(4-k)\gamma_4^2\Delta_0c^2}},
\end{eqnarray}
where ${}_2F_1(a_1,a_2;a_3;a_4)$ is the hypergeometric function \citep{Abramowitz1979}. It is not easy to get the crossing radius from above equation. However, under this case, we have $\Delta_0\gg{R}/\gamma_4^2$ and $\Delta\simeq\Delta_0$. Therefore, Equation (\ref{F1}) can be simplified to
\begin{eqnarray}\label{m33}
m_3=\left[\displaystyle\frac{16\pi EAm_p}{\displaystyle(3-k)(4-k)\gamma_4^2\Delta_0c^2}\right]^{\frac{1}{2}}R^{\frac{4-k}{2}}.
\end{eqnarray}
The RS crosses the ejecta when $m_3=E/(\gamma_4c^2)$. So the crossing radius is found to be
\begin{eqnarray}\label{Rcross1}
\begin{aligned}
R_\Delta &=\left[\displaystyle\frac{\displaystyle(3-k)(4-k)E\Delta_0}{16\pi Am_p c^2}\right]^{\frac{1}{4-k}}\\
&=\left(\frac{4-k}{4}\Delta_0{l}^{3-k}\right)^{\frac{1}{4-k}}.
\end{aligned}
\end{eqnarray}

\begin{table*}
	\centering
	\caption{Semi-analytical expressions for $R_\Delta,\,t_\Delta,\,{x}_\Delta$ and ${y}({x})$ under the condition of energy conservation.}
	\begin{tabular}{ccllccl} 
		\hline\hline
		& Expression & $f_1$ & $f_2$ & $s$ & Error\\
		\hline
 \textsc{$\displaystyle\frac{R_\Delta}{\gamma_4^2\Delta_0}$} & $\displaystyle(f_1^{s}+f_2^{s})^{1/s}$ & $\displaystyle\left[\frac{(3-k)(4-k)}{4}\xi^{2(3-k)}\right]^{\frac{1}{4-k}}$ & $\displaystyle\left[\frac{3(3-k)^2}{8}\xi^{2(3-k)}\right]^{\frac{1}{3-k}}$ & $3.24-k$ & $<7\%$ \\
 
 \textsc{$\displaystyle\frac{t_\Delta}{{\Delta_0}/{2c}}$} & $\displaystyle\left(f_1^{s}+f_2^{s}\right)^{1/s}$ & $1$ & $\displaystyle\frac{19-5k}{2(5-k)}\left[\frac{3(3-k)^2}{8}\right]^{\frac{1}{3-k}}\xi^2$ &  {$0.56-0.07k$} & $<10\%$
 \\
 
 \textsc{${y}_\Delta$} & $\displaystyle(f_1^{-s}+f_2^{-s})^{-1/s}$ & $\displaystyle\left(\frac{2}{4-k}\right)^{\frac{2-k}{2(4-k)}}\left(\frac{3-k}{4-k}\right)^{\frac{1}{2(4-k)}}\xi^{\frac{3-k}{4-k}}$ & $\displaystyle\left(\frac{4}{13-3k}\right)^{\frac{1}{2}}$ & $2$ & $<6\%$
 \\
 
 \textsc{$\displaystyle{}^{\textrm{a}}\mathcal{W}({x})$} & $\displaystyle(f_1^{s}+f_2^{s})^{1/s}$ & $\displaystyle\left(\frac{4-k}{2}\right)^{\frac{2-k}{4-k}}\left(\frac{4-k}{3-k}\right)^{\frac{1}{4-k}}\xi^{-\frac{2(3-k)}{4-k}}{x}^{\frac{2-k}{2}}$ & $\displaystyle\frac{3(3-k)}{4}{x}^{\frac{3-k}{2}}$ & $1$ & $<3\%$
 \\
\hline\hline
\end{tabular}
\label{tab:simydelta1}
\vspace{0.1cm}
\newline
$[{\textrm{A}}]$: ${y}({x})=[1+\mathcal{W}({x})]^{-1/2}$. ~~~~~~~~~~~~~~~~~~~~~~~~~~~~~~~~~~~~~~~~~~~~~~~~~~~~~~~~~~~~~~~~~~~~~~~~~~~~~~~~~~~~~~~~~~~~~~~~~~~~~~~~~~~~~~~~~~~~~~~~~~~~~~~~~~~~~~~~~~~~~~~~~~~~~~~~~~~~~~~~~~~~~~~~~~~~~~~~~~~~~~~~~~~~~~~~~~~~~~~~~~~
\end{table*}

On the contrary, in the thin-shell case of NRS case, ${y}\rightarrow{1}$ can be approximated in most of time. Equation (\ref{m31}) then can be reduced into
\begin{eqnarray}\label{m34}
m_3\textrm{d}m_3=\frac{16\pi EAm_p}{\displaystyle3(3-k){c}^2}R^{2-k}\textrm{d}R.
\end{eqnarray}
We then obtain the evolution of $m_3$ and the crossing radius as
\begin{eqnarray}\label{m35}
m_3=\left[\frac{32\pi{}EAm_p}{3(3-k)^2c^2}\right]^{\frac{1}{2}}R^{\frac{3-k}{2}}
\end{eqnarray}
and
\begin{eqnarray}\label{Rcrosydelta1}
R_\Delta =\left[\displaystyle\frac{\displaystyle3(3-k)^2E}{32\pi{}A\gamma_4^2m_pc^2}\right]^{\frac{1}{3-k}}
=\left[\displaystyle\frac{\displaystyle3(3-k)}{8\gamma_4^2}l^{3-k}\right]^{\frac{1}{3-k}}.~~
\end{eqnarray}
$R_\Delta$ can be written as the function of $\xi$
\begin{eqnarray}\label{Rcross3}
\frac{R_\Delta}{\gamma_4^2\Delta_0}=
\begin{cases}
  \displaystyle\left[\frac{(3-k)(4-k)}{4}\xi^{2(3-k)}\right]^{\frac{1}{4-k}} & (\textrm{RRS})\cr\cr
  \displaystyle\left[\frac{3(3-k)^2}{8}\xi^{2(3-k)}\right]^{\frac{1}{3-k}} & (\textrm{NRS})
  \end{cases}.~~
\end{eqnarray}

On the other hand, the energy conservation condition Equation (\ref{energy2}) can be reformed into
\begin{eqnarray}\label{chii}
\zl{{y}=\frac{\gamma_3}{\gamma_4}=\left(1+2\gamma_4\frac{m_2}{m_3}\right)^{-\frac{1}{2}},}
\end{eqnarray}
\cu{where $m_2$ is the mass of swept gas by the FS and $m_3$ in two limiting cases (RRS case and NRS case) are represented by Equations (\ref{m33}) and (\ref{m35}), respectively.}
Since both $m_2$ and $m_3$ are functions of the shock radius $R$, Equation (\ref{chii}) gives the evolution of the Lorentz factor of the shocked region. Substituting expressions of $m_2$ and $m_3$ into Equation (\ref{chii}), we have
\begin{eqnarray}\label{chiii}
{y}=
\begin{cases}
 \displaystyle\left[1+\left(\frac{l^{3-k}}{(4-k)\gamma_4^4\Delta_0}\right)^{-\frac{1}{2}}R^{\frac{2-k}{2}}\right]^{-\frac{1}{2}} & (\textrm{RRS})\cr\cr
  \displaystyle\left[1+\left(\frac{2l^{3-k}}{3(3-k)\gamma_4^2}\right)^{-\frac{1}{2}}R^{\frac{3-k}{2}}\right]^{-\frac{1}{2}} & (\textrm{NRS})
\end{cases}.~~~~
\end{eqnarray}
Note that $f$ can be written as
\begin{eqnarray}\label{fff}
f=
\begin{cases}
 \displaystyle\frac{l^{3-k}}{\displaystyle(3-k)R^{2-k}\gamma_4^2\Delta_0} & (\textrm{RRS})\cr\cr
 \displaystyle\frac{1}{3-k}\left(\frac{l}{R}\right)^{3-k} & (\textrm{NRS})
\end{cases},
\end{eqnarray}
we can get
\begin{eqnarray}\label{gamma3}
\gamma_3=y\gamma_4=
\begin{cases}
 \displaystyle\left(\frac{3-k}{4-k}\right)^{\frac{1}{4}}\gamma_4^{\frac{1}{2}}f^{\frac{1}{4}} & (\textrm{RRS})\cr\cr
 \displaystyle\gamma_4\left(1-\sqrt{\frac{21\epsilon}{16}}\right) & (\textrm{NRS})
\end{cases},
\end{eqnarray}
where $\epsilon\equiv2\gamma_4^2f^{-1}/7$, the same as the one defined in \cite{Sari1995}. Comparing with the solutions obtained by \cite{Sari1995} \zl{(in the case of $k=0$ and under the assumption of pressure balance condition)}, which is $\gamma_3=\gamma_4^{1/2}f^{1/4}/\sqrt{2}$ in the thick-shell case and $\gamma_3=\gamma_4(1-\sqrt{\epsilon})$ in the thin-shell case, one can find the difference between the results in energy conservation condition \zl{(see Equation (\ref{gamma3}))} and that in pressure equilibrium condition is only a factor of few. For example, \zl{in the case of RRS and $k=0$, the ratio of $\gamma_3$ in the assumption of energy conservation to that in the assumption of pressure balance equals $3^{1/4}$, which is roughly $1.3$}.

If we take the notion of ${x}=R/R_\Delta$ in Equation (\ref{chiii}) and substitute the expressions of $R_\Delta$ into it, we arrive at
\begin{eqnarray}\label{chi5}
{y}=
\begin{cases}
\left[1+\left(\frac{4-k}{2}\right)^{\frac{2-k}{4-k}}\left(\frac{4-k}{3-k}\right)^{\frac{1}{4-k}}\xi^{-\frac{2(3-k)}{4-k}}{x}^{\frac{2-k}{2}}\right]^{-\frac{1}{2}} & (\textrm{RRS})\cr\cr
  \displaystyle\left[1+\frac{3(3-k)}{4}{x}^{\frac{3-k}{2}}\right]^{-\frac{1}{2}} & (\textrm{NRS})
\end{cases}.
\end{eqnarray}

At the crossing radius, we have ${x}=1$ and obtain
\begin{eqnarray}\label{chi6}
{y}_\Delta=
\begin{cases}
\displaystyle\left(\frac{2}{4-k}\right)^{\frac{2-k}{2(4-k)}}\left(\frac{3-k}{4-k}\right)^{\frac{1}{2(4-k)}}\xi^{\frac{3-k}{4-k}}& (\textrm{RRS})\cr\cr
  \displaystyle\left(\frac{4}{13-3k}\right)^{\frac{1}{2}}& (\textrm{NRS})
\end{cases}.
\end{eqnarray}
Note that at the crossing radius (or ${x}=1$), ${y}$ is not very close to $1$ in the thin-shell case, implying that the RS has actually converted a non-negligible fraction of kinetic energy to thermal energy, which are often underestimated since people often take ${y}_\Delta=1$ for simplicity. A similar conclusion is found in our extended solutions for the standard FS-RS model in the previous section. On the other hand, the crossing time can be obtained by $t_\Delta=\int\textrm{d}R/(2\gamma_3^2c)=R_\Delta/(2\gamma_4^2c)\int\textrm{d}{x}/{y}^2$. Considering Equation (\ref{Rcross3}) and Equation (\ref{chi5}), we can get
\begin{eqnarray}\label{tdelta1}
t_\Delta=
\begin{cases}
\displaystyle\frac{\Delta_0}{2c} & (\textrm{RRS})\cr\cr
  \displaystyle\frac{\Delta_0}{2c}\frac{19-5k}{2(5-k)}\left[\frac{3(3-k)^2}{8}\right]^{\frac{1}{3-k}}\xi^2 & (\textrm{NRS})
\end{cases}.
\end{eqnarray}

 {Similarly, the semi-analytical solutions for $R_\Delta$ and $t_\Delta$ can be constitute with the analytical solutions in the two limiting cases, in the form of $(f_1^s+f_2^s)^{1/s}$ and ${y}_\Delta$ in the form of $(f_1^{-s}+f_2^{-s})^{-1/s}$.} We present the results in Figure \ref{fig:Fig5} and summarize the expressions in Table \ref{tab:simydelta1}. The evolution of ${y}$ versus ${x}$ can be also obtained based on Equation (\ref{chi5}). However, instead of constituting semi-analytical expressions directly from the analytical solutions in the two limiting cases, we firstly let $f_1$ and $f_2$ to be $1-1/{y}^2$, i.e., the second terms in the brackets of Equation (\ref{chi5}) of the two limiting cases, and then write $\mathcal{W}({x};\xi,k)=(f_1^{s}+f_2^{s})^{1/s}$ with $s=1$. Thus, we can get the evolution of ${y}$ versus ${x}$ as ${y}=(1+\mathcal{W})^{-1/2}$. Errors are $<5\%$ for $\xi$ from $0.01$ to $100$ and $k=0, 1, 2$,  {as shown in Figure~\ref{fig:Fig6}}.

\subsection{\zl{Application guide}}
\label{subsec: Guide}

\begin{figure}
\includegraphics[width=1\columnwidth]{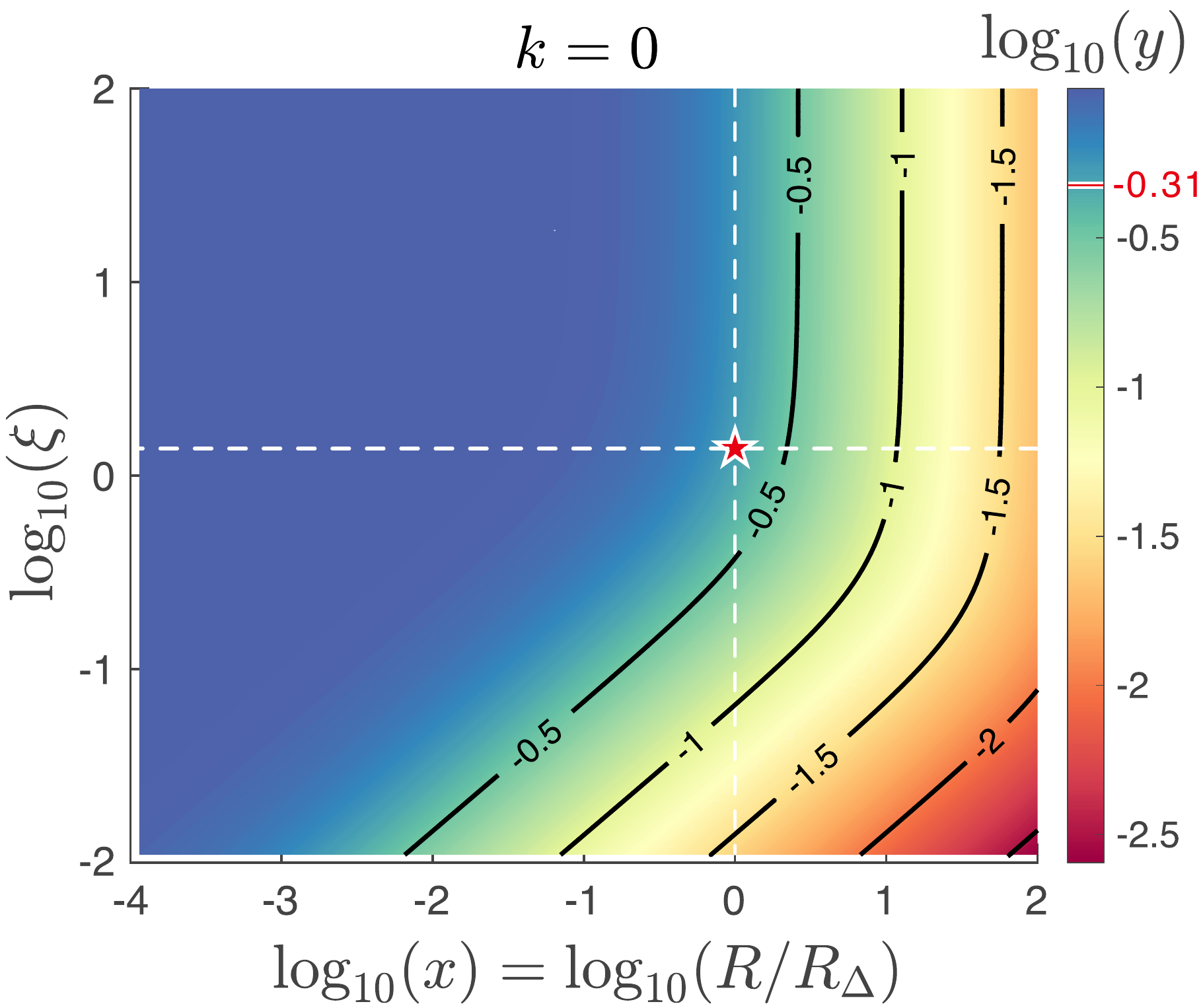}\\ 
\caption{---\cu{Contour plots: The distribution of $\log_{10}(y)$ versus $\log_{10}(\xi)$ and $\log_{10}(x)$ in the case of $k=0$. There is a red pentagram intersecting between two white dashed lines which corresponds to our example, the value of $y$ is shown in red short line on the colorbar.}} 
\label{fig:Fig7} 
\end{figure}

\cu{Now, let us give a simple example to illustrate how to use the results in Table \ref{tab:simydelta1}.
First, given a set of initial parameters: the kinetic energy of the shell $E=2\times10^{53}\, \rm erg$, shell-thickness $\Delta_0=c\times T_{90}=3\times10^{11}\, \rm cm$ with $T_{90}=10\,\rm s$ is the time span between the epochs when $5\,\%$ and $95\,\%$ of the total fluence that is collected by the detector, and the Lorentz factor $\Gamma_0=\gamma_4=300$. Here we choose ISM circumburst environment with $k=0$, and the density profile of the unshocked ambient medium $n_1=AR^{-k}=n_0(R/R_0)^{-k}$ (where $n_1=n_0$ for $k=0$ and $R_0$ is normalization factor for the characteristic length of shock). Assuming $n_0=1\,\rm cm^{-3}$ and
adopting Equations (\ref{Sedov}) and (\ref{ksi}) to find the Sedov length $l\equiv[3E/(4\pi n_0 m_p c^2)]^{1/3}=3.2\times10^{18}\,\rm cm$ and $\xi=[E/(4\pi n_0 m_p c^2\Delta_0^{3}\gamma_4^{8})]^{1/6}=1.35$, then we can plug all the related parameters into the results shown in Table \ref{tab:simydelta1} to find the crossing radius $R_\Delta=8.2\times10^{16}\, \rm cm$, crossing time $t_\Delta=47\,\rm s$, and Lorentz factor $\gamma_3=\gamma_4 y=300\times10^{-0.31}=147$ (as shown in Figure \ref{fig:Fig7}) in the shocked shell. Further applications of these equations, for example, to give parameters relevant for reverse shock emission, such as the particle density in region 3 will be shown in our future work.}

\section{Conclusions}
\label{sec: conclusion}

To summarize, we revisited the FS-RS model for early afterglows of GRBs in this paper, and tried to obtain simple semi-analytical solutions that can be easily applied in future studies. We first adopted the standard FS-RS model for GRB early afterglow and extended it to more general cases. Assuming the density profile of circumburst environment to be a power-law form of $R^{-k}$, we obtain analytical solutions of the RS dynamics in the two limiting cases, based on which we further constituted a simple semi-analytical solution for the general case. The semi-analytical solution can be reduced to the the analytical solutions in the two limiting cases if the RS is ultra-relativistic ($\xi\ll{1} $ or ${y}\simeq{0}$ or $\Delta_0\gg{R}_\Delta/\gamma_4^2$) or Newtonian ($\xi\gg{1} $ or ${y}\simeq{1}$ or $\Delta_0\ll{R}_\Delta/\gamma_4^2$), and agreed well with the numerical results of the intermediate case or the mildly-relativistic RS case. Secondly, we reconsider the hydrodynamic evolution of FS-RS system on the condition of energy conservation. Crossing radius, crossing time and the Lorentz factor of shocked material at crossing time are obtained analytically in the two limiting cases. 

The evolution of the Lorentz factors of shocked material in the entire crossing process are obtained analytically as well. The obtained quantities in energy conservation condition are only a factor of few different from those in the standard FS-RS model, except quantities at crossing time in thin-shell case. Our solutions show RS can convert a non-negligible fraction of kinetic energy of GRB ejecta into thermal energy of shocked particles in the thin-shell case, and this could make some difference in the predicted broadband afterglow light curves. Our semi-analytical expressions provide general solutions to the FS-RS evolution and may help to understand the GRBs ejecta and circumburst environment better when being applied to fit lightcurves of early afterglow, which is to be studied in the near future.

Lastly, it is worth noting that we do not consider magnetic fields in the evolution of the FS-RS. In reality, the influence of the magnetic field may not be negligible in certain conditions. Observations and theoretical modelling of GRB early afterglow (e.g., \citealt{Zhang2003}; \citealt{Fan2004}; \citealt{Troja2017}) indicate that at least some of GRB ejecta are magnetically dominated. \cu{In addition, the detection of polarization of an early GRB optical afterglow supports large scale structured magnetic fields (e.g., \citealt{Steele2009} and \citealt{Mundell2013}).
\citet{Harrison2013} also found that the magnetic fields in the GRB outflow are stronger than previously believed.}
Therefore, it may be interesting to study and discuss the behaviors of the magnetic components of the FS-RS system in the same framework in our future research.

\section{Acknowledgments}

\cu{We acknowledge the anonymous referee for the valuable suggestions.}
This work is supported by the National Key R \& D program of China under the Grant No. 2018YFA0404203, the National Natural Science Foundation of China (grant numbers 12121003, \zl{11725314,} 12041306, 11903019, U2031105), China Manned Spaced Project (CMS-CSST-2021-B11)
and Major Science and Technology Project of Qinghai Province (2019-ZJ-A10).

\section{Data availability}
The code and data underlying this article will be shared on reasonable request
to the first corresponding author.






\appendix

\section{The accuracy of numerical calculations}
\label{sec: numerical}

\begin{figure*}
\centering
\includegraphics[width=2\columnwidth]{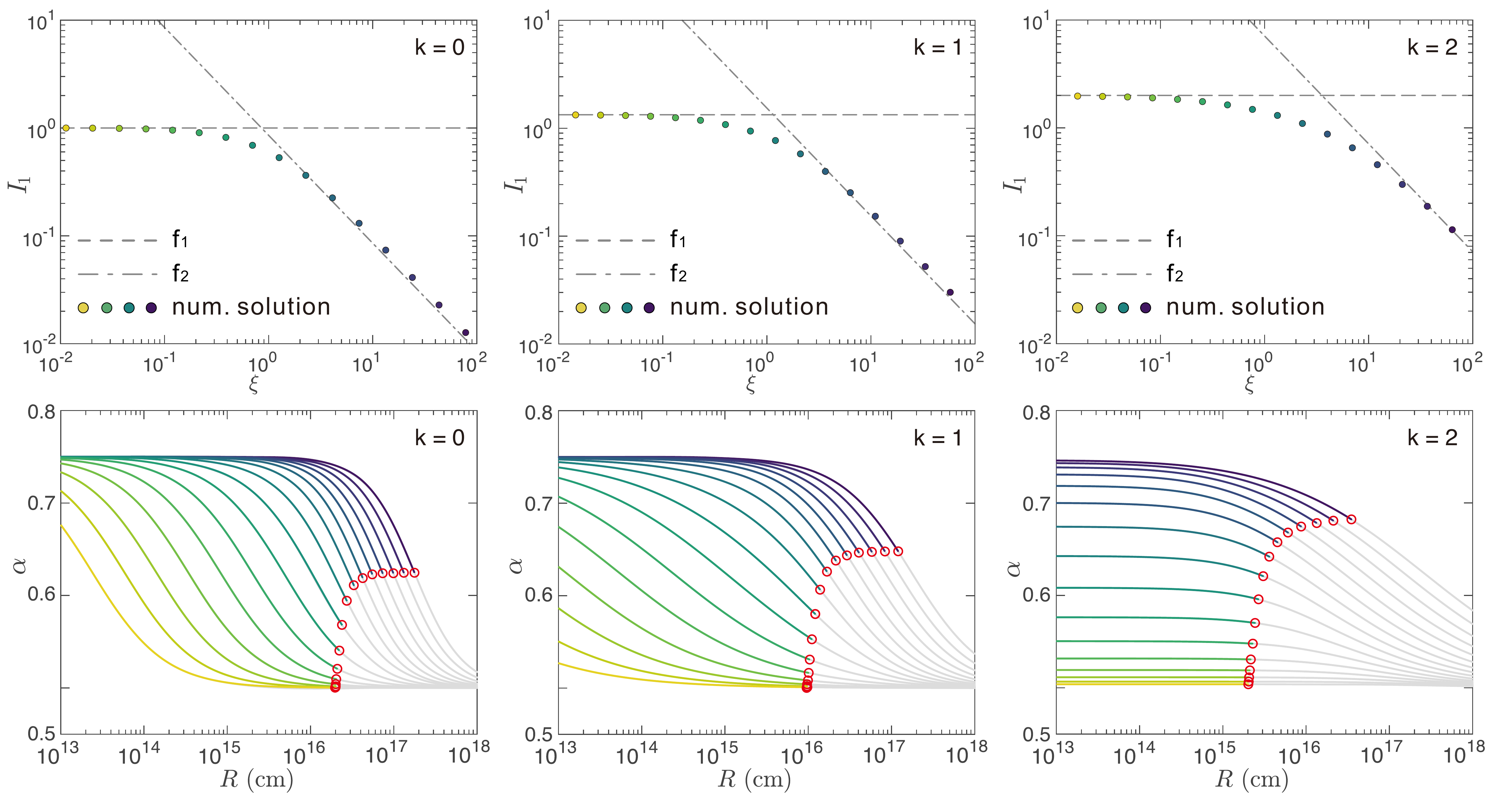}\\ 
\caption{--- The points with colors from purple (approaching to the NRS limit) to yellow (approaching to the RRS limit) in the upper panels are corresponding to the colorful curves in the lower panels which are in the same colors. The red circles attached to each curve represent the crossing radius $R_{\Delta}$ for the corresponding $\xi$.} 
\label{fig:Appendix} 
\end{figure*}

From Figures \ref{fig:Fig2} and \ref{fig:Fig3}, we notice that , in the case of NRS limit ($\xi\gg1$), the numerical solutions and the analytical solutions of the parameters, i.e., $I_1$, $I_2$ and $t_{\Delta}$ are not exactly the same, these deviations come from the the validity of the approximate method we adopt in Equation (\ref{ff1}). It can be shown in the lower panels of Figure \ref{fig:Appendix}. When $\xi\gg1$ (purple curves), the curves descend quickly around the crossing radius, while approximately to $0.5$ around the crossing radius when $\xi\ll{1}$ (yellow curves).  The value of $\alpha$ versus $R$ is partly less than $3/4$ in the thin-shell case, especially when $k=0, 1$. We know $I_1$ and $I_2$ are inversely proportional to $\alpha$, \cu{see Equation (\ref{alpha}) in the main text,} so the numerical results will be little larger than the analytical solutions. Then the contribution of the deficits to the integral in Equations (\ref{Pi1}) and (\ref{Pi2}) will lead to the deviations between the numerical solutions and the analytical solutions.

\section{\zl{Contour plots of $\log_{10}(y)$ in the cases of $k=1$ and $k=2$}}
\label{sec: contour}

\zl{Here we show the contour plots of the distribution of $\log_{10}(y)$ versus $\log_{10}(\xi)$ and $\log_{10}(x)$ in the cases of $k=1$ and $k=2$.}

\begin{figure*}
\centering
\includegraphics[width=2\columnwidth]{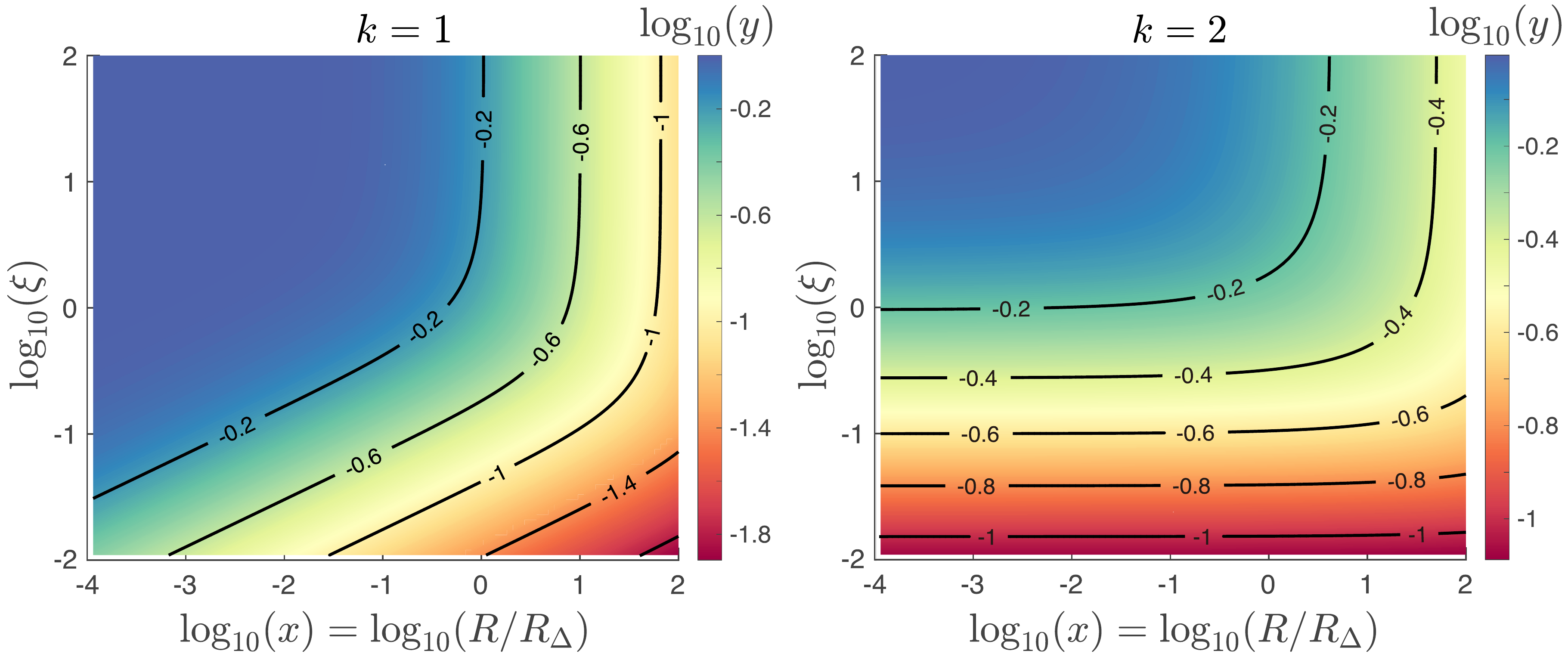}\\ 
\caption{---\cu{Contour plots: The distribution of $\log_{10}(y)$ versus $\log_{10}(\xi)$ and $\log_{10}(x)$. The panels from left to right are correspond to $k=1$ and $2$, separately.}} 
\label{fig:Appendix2} 
\end{figure*}


\bsp	
\label{lastpage}
\end{document}